\documentclass[superscriptaddress, aps, prb, twocolumn, showpacs]{revtex4}

\usepackage[pdftex]{graphicx}
\usepackage{epstopdf}

\begin{document}

\title{Doping evolution and polar surface reconstruction of the infinite-layer cuprate Sr$_{1-x}$La$_{x}$CuO$_{2}$}

\author{John W. Harter}
\altaffiliation[Current affiliation: ]{Department of Physics, California Institute of Technology, Pasadena, California 91125, USA}
\affiliation{Laboratory of Atomic and Solid State Physics, Department of Physics, Cornell University, Ithaca, New York 14853, USA}

\author{Luigi Maritato}
\affiliation{Department of Materials Science and Engineering, Cornell University, Ithaca, New York 14853, USA}
\affiliation{Dipartimento di Ingegneria dell'Informazione, Ingegneria Elettrica e Matematica Applicata-DIEM, University of Salerno and CNR-SPIN, 84084 Fisciano, Salerno, Italy}

\author{Daniel E. Shai}
\author{Eric J. Monkman}
\affiliation{Laboratory of Atomic and Solid State Physics, Department of Physics, Cornell University, Ithaca, New York 14853, USA}

\author{Yuefeng Nie}
\affiliation{Laboratory of Atomic and Solid State Physics, Department of Physics, Cornell University, Ithaca, New York 14853, USA}
\affiliation{Department of Materials Science and Engineering, Cornell University, Ithaca, New York 14853, USA}

\author{Darrell G. Schlom}  
\affiliation{Department of Materials Science and Engineering, Cornell University, Ithaca, New York 14853, USA}
\affiliation{Kavli Institute at Cornell for Nanoscale Science, Ithaca, New York 14853, USA}

\author{Kyle M. Shen}
\email[Author to whom correspondence should be addressed: ]{kmshen@cornell.edu}
\affiliation{Laboratory of Atomic and Solid State Physics, Department of Physics, Cornell University, Ithaca, New York 14853, USA}
\affiliation{Kavli Institute at Cornell for Nanoscale Science, Ithaca, New York 14853, USA}

\date{\today}

\begin{abstract}
We use angle-resolved photoemission spectroscopy to study the doping evolution of infinite-layer Sr$_{1-x}$La$_{x}$CuO$_{2}$ thin films grown by molecular-beam epitaxy. At low doping, the material exhibits a dispersive lower Hubbard band typical of the superconducting cuprate parent compounds. As carriers are added to the system, a continuous evolution from charge-transfer insulator to superconductor is observed, with the initial lower Hubbard band pinned well below the Fermi level and the development of a coherent low-energy band with electron doping. This two-component spectral function emphasizes the important role that strong local correlations play even at relatively high doping levels. Electron diffraction probes reveal a ${p(2\times2)}$ surface reconstruction of the material at low doping levels. Using a number of simple assumptions, we develop a model of this reconstruction based on the polar nature of the infinite-layer structure. Finally, we provide evidence for a thickness-controlled transition in ultrathin films of SrCuO$_2$ grown on nonpolar SrTiO$_3$, highlighting the diverse structural changes that can occur in polar complex oxide thin films.
\end{abstract}


\maketitle

\section{Introduction}

The generic aspects of the hole-doped side of the cuprate phase diagram have been firmly established by investigating a multitude of hole-doped material families. In contrast, there exist only two families from which most of our understanding of electron doping in the cuprates is derived: \textit{Re}$_{2-x}$Ce$_{x}$CuO$_{4}$, where \textit{Re} is a rare earth element, and ``infinite-layer'' Sr$_{1-x}$La$_x$CuO$_2$ \cite{armitage2010}. One conspicuous difference between electron and hole doping is the strength of $(\pi,\pi)$ antiferromagnetic order, which can persist up to an electron doping of $x = 0.14$ \cite{luke1990} and may even coexist with superconductivity in some materials \cite{yamada2003,alff2003,harter2012}. In all of the hole-doped cuprates, on the other hand, antiferromagnetism is rapidly suppressed by $x \approx 0.03$ and does not coexist with superconductivity. These facts highlight a clear asymmetry in the doping phase diagram of the cuprates, with important ramifications for theories of high-temperature superconductivity. Sr$_{1-x}$La$_x$CuO$_2$ is an ideal material for studying electron doping in the cuprates because of its uncomplicated structure, which consists of perfectly square and flat CuO$_2$ planes separated by simple Sr$_{1-x}$La$_x$ charge reservoir layers.

Bulk growth of Sr$_{1-x}$La$_x$CuO$_2$ is limited to polycrystalline samples due to the required high growth pressures. High quality thin films, however, can be grown by taking advantage of epitaxial stabilization \cite{tsukahara1991,norton1994,karimoto2001,karimoto2004,leca2006}. Indeed, within the last decade, complex oxide thin films have attracted increasing attention for the diverse electronic systems that they host \cite{heber2009,mannhart2010}. Polar structures with alternating charged atomic planes are particularly prevalent among the array of oxide structures commonly studied \cite{ohtomo2004,nakagawa2006,reyren2007}.  As thin films of these materials are grown, this alternating polarity leads to a thermodynamically unstable electric potential divergence---a so-called ``polar catastrophe''---which is prevented by a structural or electronic reconstruction. Undoped SrCuO$_2$, consisting of alternating Sr$^{2+}$ and (CuO$_2$)$^{2-}$ planes and no mobile charges to screen the electrostatic potential buildup, is such an example.

In this paper we use \textit{in situ} angle-resolved photoemission spectroscopy (ARPES) and electron diffraction to study both the doping evolution as well as the polar reconstruction in epitaxially-stabilized Sr$_{1-x}$La$_{x}$CuO$_{2}$ thin films grown by molecular-beam epitaxy. In Sec.\ \ref{sectionMethods}, we describe the film growth and experimental details. In Sec.\ \ref{sectionUndoped}, we present measurements of Sr$_{1-x}$La$_x$CuO$_2$ at low doping levels, showing a dispersive lower Hubbard band (LHB) characteristic of other parent cuprates. Section \ref{sectionDoping} demonstrates that with increased electron doping, a continuous evolution from insulator to superconductor occurs as spectral weight fills in the charge-transfer gap. In Sec.\ \ref{sectionReconstruction}, we describe electron diffraction probes that show evidence of a surface reconstruction consistent with the mitigation of a polar catastrophe in Sr$_{1-x}$La$_{x}$CuO$_{2}$. We also present evidence supporting a recent theoretical prediction of a thickness-controlled structural transition in ultrathin films of SrCuO$_2$ grown on nonpolar SrTiO$_3$. Finally, Sec.\ \ref{sectionConclusion} offers conclusions and implications of our work.

\begin{figure*}
\includegraphics{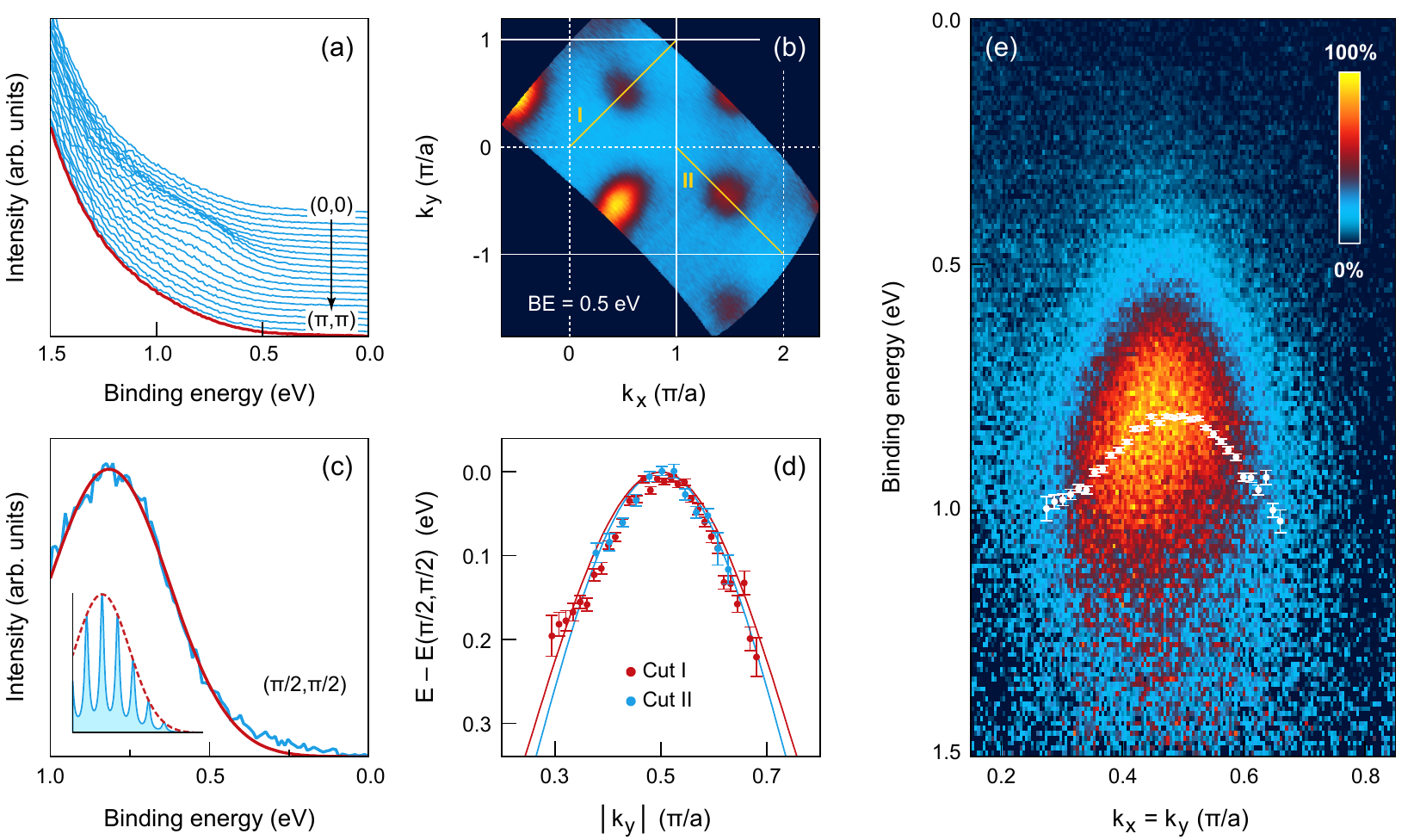}
\caption{\label{figure1}{The low-energy electronic structure of Sr$_{0.99}$La$_{0.01}$CuO$_{2}$.  (a) Energy distribution curves (EDCs), offset for clarity, along a diagonal cut from $(0,0)$ to $(\pi,\pi)$ through the LHB, which is visible as a bump at the foot of the valence band. The bold red line shows the background EDC subtracted from the data in the remaining panels in order to enhance the LHB feature.  (b) Momentum space map of spectral weight at a binding energy of 0.5 eV showing the LHB at $(\pi/2,\pi/2)$ and equivalent points.  (c) EDC at $(\pi/2,\pi/2)$ after background subtraction. The peak has a Franck-Condon lineshape (demonstrated schematically in the lower left inset) and can be fit to a Gaussian (red curve). The ``foot'' in the low-binding-energy region deviates slightly from the Gaussian and most likely reflects low-energy levels occupied by the small amount of dopants added to the sample.  (d) Dispersion of the LHB as a function of momentum along Cut I from $(0,0)$ to $(\pi,\pi)$ (red points) and along perpendicular Cut II from $(\pi,0)$ to $(2\pi,-\pi)$ (blue points), as determined by Gaussian fitting. The smooth curves show the dispersion predicted by the $t$-$t'$-$t''$-$J$ model with $J = 150$ meV.  (e) Experimental spectrum along Cut I after background subtraction. The white dots reproduce the dispersion shown in panel (d).}}
\end{figure*}

\section{Experimental details}
\label{sectionMethods}

Sr$_{1-x}$La$_{x}$CuO$_{2}$ thin films ($x$ = 0, 0.01, 0.05, and 0.10) with a thickness of 60 unit cells (20 nm) and terminated with CuO$_{2}$ were deposited using a Veeco GEN10 dual-chamber oxide molecular-beam epitaxy system. Films were grown epitaxially on (110) GdScO$_{3}$ substrates, which have a distorted perovskite structure with a pseudocubic lattice constant of 3.968 \AA \cite{uecker2008}. Shuttered layer-by-layer deposition was performed in a background of 80\% pure distilled O$_{3}$ at a pressure of $1\times10^{-6}$ Torr and with a substrate growth temperature of 510 $^\circ$C. Depositions were monitored using reflection high-energy electron diffraction (RHEED). After growth, samples were reduced by vacuum annealing at 520 $^\circ$C for 30 minutes in order to eliminate excess oxygen atoms. The films were then cooled to 200 $^\circ$C before immediate transfer under ultra-high vacuum into the ARPES chamber. Samples with $x = 0.10$ were superconducting, exhibiting bulk resistance transitions in the range 25 $\pm$ 5 K. Further details of the film growth can be found in Ref.\ \citenum{maritato2013}.

ARPES measurements were performed with a VG Scienta R4000 electron spectrometer and He-I$\alpha$ photons (21.2 eV) at a base pressure of $7\times10^{-11}$ Torr and with an instrumental resolution better than $\Delta E$ = 20 meV and $\Delta k$ = 0.03 \AA$^{-1}$. The sample temperature was held at 200, 30, and 10 K for $x$ = 0.01, 0.05, and 0.10, respectively. The Fermi level $E_{\mathrm{F}}$ was determined by measuring polycrystalline gold in electrical contact with the sample. Experimental results were confirmed by studying multiple samples. After ARPES measurements, samples were characterized by \textit{in situ} low-energy electron diffraction (LEED) to examine surface structure and quality.  X-ray photoelectron spectroscopy was used to verify the stoichiometry of the films, measuring a difference of less than 0.01 between the measured and nominal lanthanum doping level $x$.

\begin{figure*}
\includegraphics{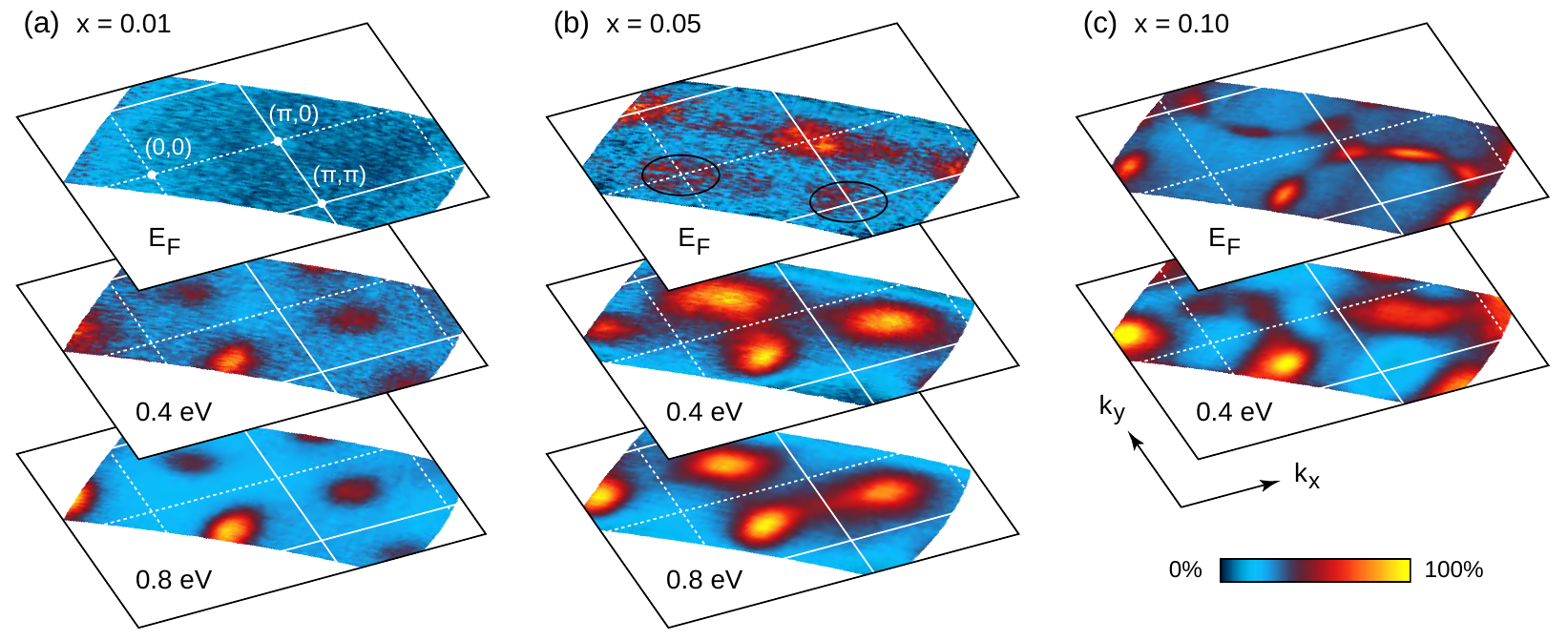}
\caption{\label{figure2}{Momentum space evolution of spectral intensity with doping. Constant energy spectral maps for (a) $x = 0.01$, (b) $x = 0.05$, and (c) $x = 0.10$, integrated within $\pm$50 meV of the specified binding energy. At the Fermi level, the insulating $x = 0.01$ sample shows no spectral weight, while the $x = 0.05$ sample shows an accumulation of weight at $(\pi,0)$. By $x = 0.10$, the electron pocket at $(\pi,0)$ is well established and additional spectral weight is apparent at $(\pi/2,\pi/2)$. This weight is due to the finite integration region of the map rather than a true band crossing at the Fermi level \cite{harter2012}. At higher binding energies, there is clear evidence for a coexisting LHB with intensity near $(\pi/2,\pi/2)$ for all three doping levels. The circled regions at the top of panel (b) show shadow band reflections from spectral weight at $(\pi,0)$ due to a $p(2\times2)$ surface reconstruction in this sample, as discussed in Sec.\ \ref{sectionReconstruction}.}}
\end{figure*}

\section{Parent electronic structure}
\label{sectionUndoped}

The undoped parent compounds of the cuprates are charge-transfer insulators in which strong local Coulomb interactions dominate over a conventional band structure picture. Instead, the low-energy electronic structure is composed of a LHB, which typically has a bandwidth of $\sim$0.3 eV and a maximum at $(\pi/2,\pi/2)$ \cite{dagotto1994}. In Fig.\ \ref{figure1}, we show ARPES data for Sr$_{0.99}$La$_{0.01}$CuO$_{2}$, where $x = 0.01$ was intentionally added to prevent electrostatic charging of the sample (observed in stoichiometric SrCuO$_{2}$ films), pinning the chemical potential near the bottom of the upper Hubbard band. A small shoulder in the tail of the valence band is clearly present. After subtraction of a background EDC obtained by averaging the valence band tail near $(0,0)$, we observe a dispersive peak with a broad lineshape, characteristic of the LHB in other parent cuprates \cite{shen2004,armitage2002}. The spectral shape of the LHB is due to Franck-Condon broadening in which the coupling to a bosonic mode causes the spectral function to split into a set of discrete peaks. Each peak represents a resonance with a different boson occupation number, with the true quasiparticle pole residing in the low-binding-energy tail of intensity \cite{shen2004}. At $(\pi/2,\pi/2)$, the lineshape can be well-fit to a simple three-parameter Gaussian function, shown in Fig.\ \ref{figure1}(c), with the intensity maximum in the LHB at a binding energy of 0.81 eV and a full-width at half-maximum (FWHM) of 0.41 eV. This compares well with the LHB observed in other cuprate parent compounds, such as Ca$_2$CuO$_2$Cl$_2$ \cite{shen2004} and Nd$_2$CuO$_4$ \cite{armitage2002}, where the FWHM of the LHB is measured to be 0.34 eV and 0.36 eV, respectively.

As Fig.\ \ref{figure1}(d) shows, the LHB is dispersive, exhibiting a symmetric energy maximum at $(\pi/2,\pi/2)$. The $t$-$t'$-$t''$-$J$ model is often used to describe the motion of a doped carrier in a two-dimensional antiferromagnetic insulator, with the $t'$ and $t''$ parameters necessary to reproduce the clearly observed dispersion in the $(\pi,0)$ to $(0,\pi)$ direction. By fixing $J = 150$ meV and allowing $t'$ and $t''$ to vary in order to match the experimentally measured curvature along the $(0,0)$ to $(\pi,\pi)$ and transverse directions, we obtain a good fit to the data with $t' = -53$ meV and $t'' = 66$ meV. This is similar to the values $J = 140$ meV, $t' = -38$ meV, and $t'' = 22$ meV from Ref.\ \citenum{tohyama2000}, obtained by fitting to self-consistent Born approximation calculations of Sr$_2$CuO$_2$Cl$_2$, another undoped cuprate. The similarity suggests a universality of the electronic structure of the cuprates in the limit of low doping. While weakly interacting formalisms such as density functional theory predict a metallic state, the $t$-$t'$-$t''$-$J$ model in conjunction with Franck-Condon broadening can describe the observed electronic structure, accounting for the bandwidth change from $8t \sim 3$ eV to $2J \sim 0.3$ eV and the dispersion symmetry around $(\pi/2,\pi/2)$.

\section{Doping evolution of low-energy states}
\label{sectionDoping}

\begin{figure*}
\includegraphics{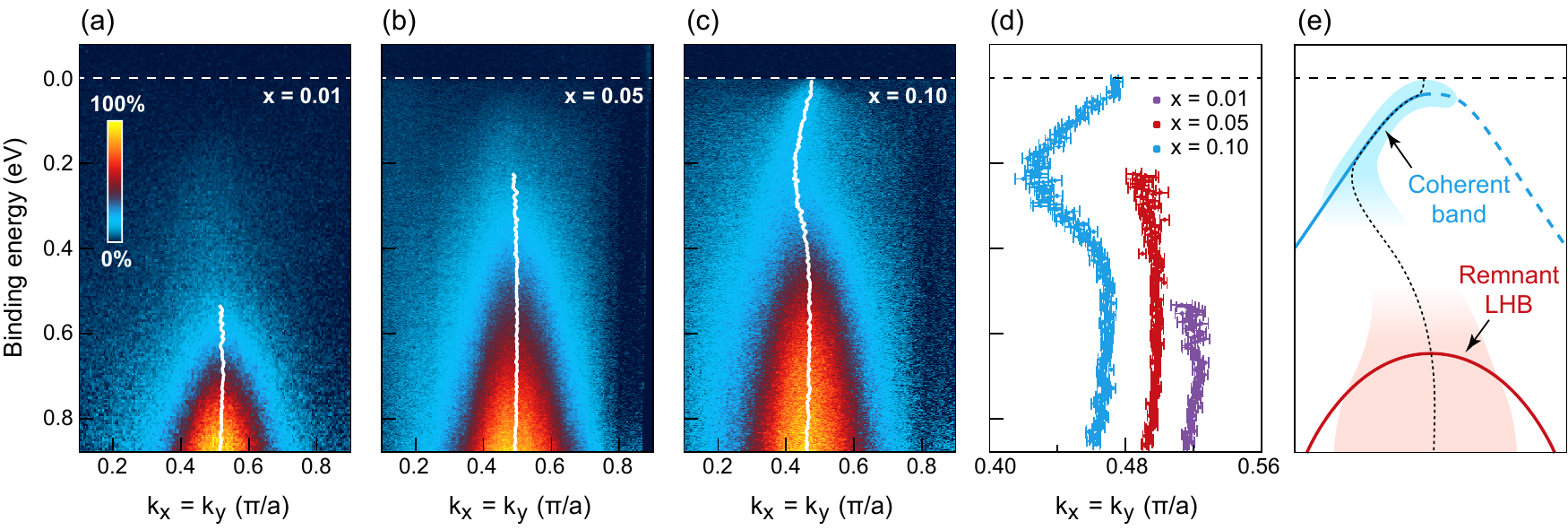}
\caption{\label{figure3}{Evolution of electronic structure with doping. Spectra along $(0,0)$ to $(\pi,\pi)$ after background subtraction (a) for $x = 0.01$, (b) for $x = 0.05$, and (c) for $x = 0.10$. As the doping level increases, the spectral weight of the LHB shifts to lower binding energy and gradually fills in the charge-transfer gap. At $x = 0.10$, a coherent band near the Fermi level is visible.   (d) Momentum distribution curve (MDC)-derived dispersion for $x =$ 0.01, 0.05, and 0.10 (also shown as white lines in the preceding panels). The LHB maximum appears to shift away from $(\pi/2,\pi/2)$ with increased doping. The ``boomerang'' phenomenon is clearly visible for $x = 0.10$.  (e) Schematic diagram showing the qualitative form of the spectral function for $x = 0.10$. Spectral weight fills in the charge-transfer gap, forming a coherent band on top of the remnant LHB. The ``boomerang'' phenomenon in MDC-derived dispersions, shown by the black dashed line, is an artifact arising from the presence of two bands.}}
\end{figure*}

\begin{figure}
\includegraphics{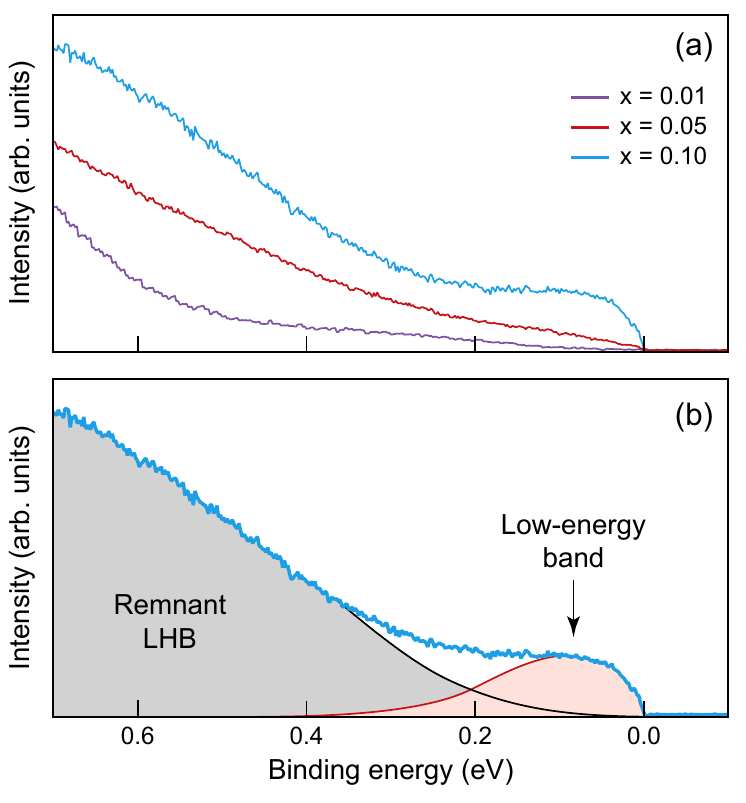}
\caption{\label{figure4}{Energy distribution curves.  (a) Doping dependence of EDCs at $(\pi/2,\pi/2)$ for $x =$ 0.01, 0.05, and 0.10. As the doping level increases, states are filled in near the Fermi level.  (b) An illustration of the two-component spectral function for $x = 0.10$. The EDC is made up of a coherent low-energy band and an incoherent high-energy remnant LHB.}}
\end{figure}

Upon addition of electrons into the CuO$_2$ planes via the substitution of trivalent lanthanum for divalent strontium in the intervening layers, the upper Hubbard band in Sr$_{1-x}$La$_{x}$CuO$_{2}$ evolves into a metallic Fermi surface with spectral weight first appearing at $(\pi,0)$ and equivalent points, as evidenced by the Fermi surface map of the $x = 0.05$ sample in Fig.\ \ref{figure2}(b). This behavior is identical to that seen in the electron-doped \textit{Re}$_{2-x}$Ce$_{x}$CuO$_{4}$ family \cite{armitage2002}. At $x = 0.10$, the Fermi surface consists solely of an electron pocket centered at $(\pi,0)$, although spectral weight is still apparent at $(\pi/2,\pi/2)$. As Ref.\ \citenum{harter2012} describes, this unusual Fermi surface is due to strong $(\pi,\pi)$ antiferromagnetism, which gaps the quasiparticles near $(\pi/2,\pi/2)$. Recent studies of the \textit{Re}$_{2-x}$Ce$_{x}$CuO$_{4}$ family have shown a clear link between oxygen content and the gapping of portions of the Fermi surface by antiferromagnetism \cite{richard2007,richard2008,song2012}. Chemical and structural similarities between the two families strongly suggests that the energy pseudogap near $(\pi/2,\pi/2)$ observed for Sr$_{0.90}$La$_{0.10}$CuO$_{2}$ may also be due to incomplete oxygen reduction. A further discussion of the oxygen reduction step can be found in Sec.\ \ref{sectionReconstruction}. Figure \ref{figure2} also shows maps at higher binding energies, where all three doping levels show evidence of a remnant LHB which takes the form of diffuse regions of spectral weight centered near $(\pi/2,\pi/2)$ and coexistent with the coherent bands dispersing through the Fermi level.

As Fig.\ \ref{figure3} shows, along $(0,0)$ to $(\pi,\pi)$ we observe that rather than closing abruptly, the charge-transfer gap gradually fills with spectral weight upon electron doping from 1\% to 10\%. This is accompanied by a qualitative change in the nature of the low-energy excitations of the system: the localized states of the LHB give way to a coherent itinerant band dispersing through the Fermi level. Figures \ref{figure3}(c,d) highlight a notable feature in the dispersion derived from a MDC analysis: at approximately 0.2 eV, the dispersion appears to ``boomerang'' backwards. This phenomenon is observed at multiple equivalent points in momentum space and is inconsistent with single-band physics. Instead, the effect likely arises from a two-component spectral function illustrated in Fig.\ \ref{figure4}: a coherent low-energy band forming the Fermi surface and a large contribution of incoherent spectral weight at higher energies derived from a remnant LHB that survives even at $x = 0.10$. This behavior differs from the hole-doped cuprates or the \textit{Re}$_{2-x}$Ce$_{x}$CuO$_{4}$ family, where so-called ``waterfalls'' are observed at higher binding energies \cite{graf2007,inosov2007,ikeda2009,moritz2009}. Recent theoretical calculations that take into account strong electron correlations have predicted such a coexistence of a low-energy band and an incoherent high-energy branch at 10\% electron doping \cite{senechal2005,weber2010}. Figure \ref{figure3}(d) also shows that the position of the LHB maximum appears to shift from $(\pi/2,\pi/2)$ towards $(0,0)$ with doping at a rate of approximately $8.5\times10^{-3}$ $(\pi/a)$/\%, which is quantitatively similar to the behavior observed for hole-doped Ca$_{2-x}$Na$_{x}$CuO$_{2}$Cl$_{2}$, where the LHB shifts with doping at a rate of $7.5\times10^{-3}$ $(\pi/a)$/\% \cite{shen2004}. Interestingly, the maximum shifts in the same direction for both electron and hole doping, counter to what would be expected from band structure calculations. The prominence of the remnant LHB in the experimental data highlights the important role that strong local electron correlations play in the electronic structure of Sr$_{1-x}$La$_{x}$CuO$_{2}$ even at relatively high doping levels.

We observe an interesting feature in the Fermi surface map of the $x = 0.05$ sample, as circled in Fig.\ \ref{figure2}(b): weak but finite spectral weight at $(0,0)$ and $(\pi,\pi)$. Bands at these locations in momentum space are not expected by tight-binding, density functional theory, or the $t$-$t'$-$t''$-$J$ model. Instead, it appears that the observed intensity is the result of a $p(2\times2)$ surface reconstruction of the sample, which causes shadows of the real spectral weight at $(\pi,0)$ to be reflected onto these locations in momentum space. At higher binding energies, evidence of this reconstruction is absent because reflections from intense regions of the remnant LHB fall onto each other. This surface reconstruction is described further in the next section.

\section{Polar surface reconstruction}
\label{sectionReconstruction}

\begin{figure}
\includegraphics{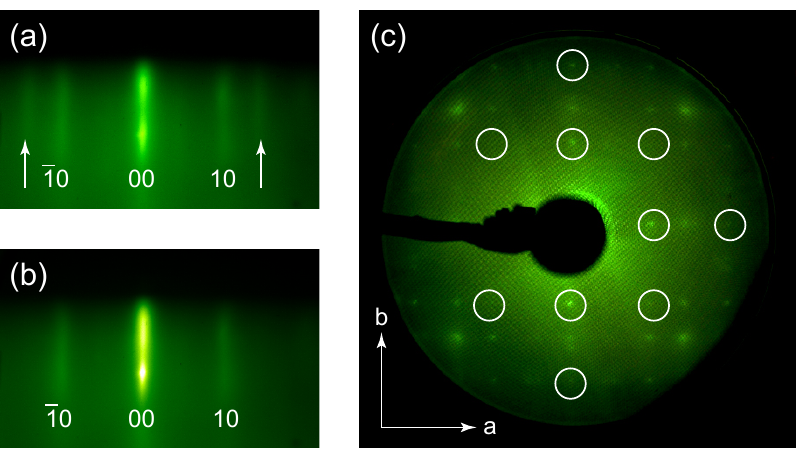}
\caption{\label{figure5}{Structural change induced by the oxygen reduction step.  (a) RHEED image along the [100]$_p$ azimuth before the vacuum annealing step for an $x = 0.10$ film. White arrows highlight extra diffraction streaks present in all as-grown films.  (b) RHEED image after oxygen reduction for the same film. The extra RHEED streaks vanish during the annealing step.  (c) LEED image of an unannealed $x = 0.10$ film taken with 100 eV electrons. Circles show where Bragg peaks are located for films with the proper infinite-layer structure; some Bragg peaks are missing while a number of incommensurate peaks are visible.}}
\end{figure}

The metastable infinite-layer structure, which can only be grown in polycrystalline form under high pressure or as an epitaxially-stabilized thin film, lies near a manifold of other structural phases, such as the edge-sharing chain-type structure \cite{gambardella1992}. One consequence is an elevated sensitivity to oxygen stoichiometry.  For example, all as-grown Sr$_{1-x}$La$_{x}$CuO$_{2}$ films require a vacuum annealing step in order to eliminate excess oxygen and form the infinite-layer structure. This feature is shared by both families of electron-doped cuprates and is generally believed to be related to the absence of apical oxygen atoms in their respective crystal structures. As Fig.\ \ref{figure5} shows, RHEED patterns before and after the oxygen reduction step show a marked structural change, and LEED performed on an unreduced film shows a number of extra diffraction peaks incommensurate with the tetragonal infinite-layer reciprocal lattice.

\begin{figure}
\includegraphics{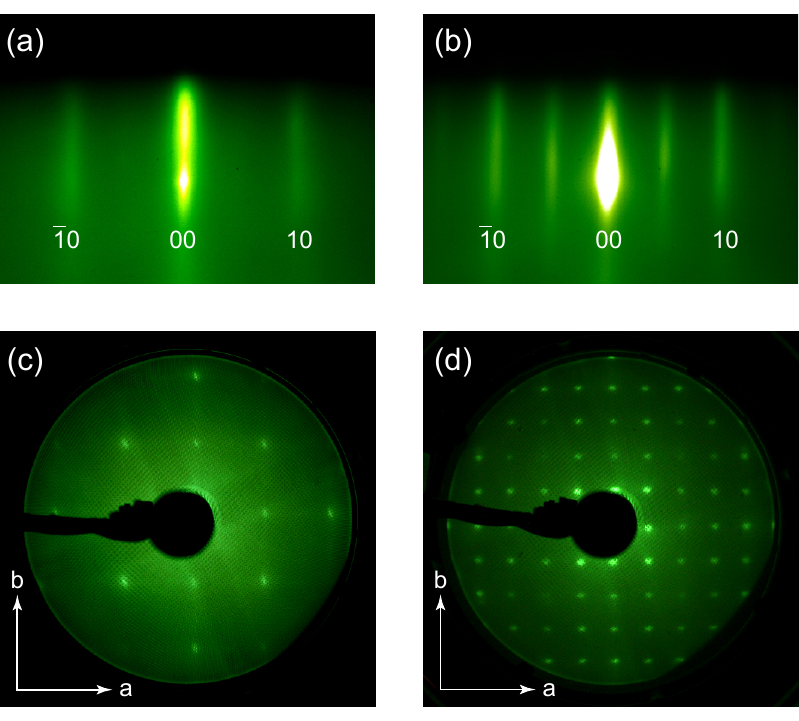}
\caption{\label{figure6}{Evidence of surface reconstruction by electron diffraction.  RHEED images along the [100]$_p$ azimuth after the vacuum annealing step for Sr$_{1-x}$La$_{x}$CuO$_{2}$ films with (a) $x = 0.10$ and (b) $x = 0$. The latter image shows extra diffraction streaks consistent with a doubled lattice constant.  LEED images taken with 100 eV electrons for films with (c) $x = 0.10$ and (d) $x = 0.05$. The latter image again shows very clear evidence of a $p(2\times2)$ surface reconstruction causing a doubling of the unit cell in both the $a$ and $b$ directions.}}
\end{figure}

Even within the correct bulk structural phase, many nominally tetragonal transition metal oxides, such as Sr$_2$RuO$_4$ \cite{damascelli2000} and SrTiO$_3$ \cite{erdman2002}, are known to support surface reconstructions because of their complex surface chemistry. Electron diffraction probes are sensitive to such reconstructions. As demonstrated in Fig.\ \ref{figure6}, both RHEED, performed after growth at high temperature, and LEED, performed at low temperature, indicate that some Sr$_{1-x}$La$_{x}$CuO$_{2}$ samples with low doping levels show a $p(2\times2)$ surface reconstruction not observed by bulk x-ray diffraction. Within the ionic limit at low doping, the infinite-layer structure is intrinsically polar, alternating between charged Sr$^{2+}$ and (CuO$_2$)$^{2-}$ layers. This introduces a thermodynamic instability towards reconstruction as films of the material are grown. We argue below that the observed $p(2\times2)$ reconstruction is most likely a result of the polar surface of Sr$_{1-x}$La$_{x}$CuO$_{2}$, and that a polar catastrophe is avoided in this material by the formation of ordered oxygen vacancies on the topmost CuO$_2$ plane.

\begin{figure}
\includegraphics{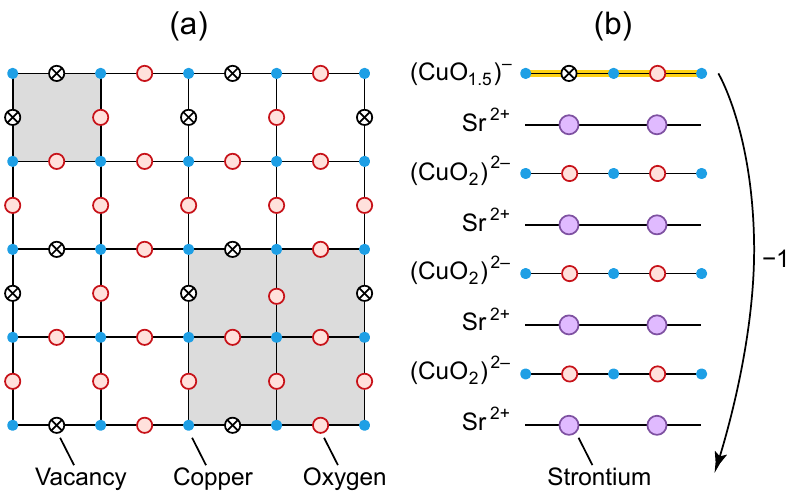}
\caption{\label{figure7}{Proposed model of surface reconstruction.  (a) One orientation of the surface structure composed of oxygen vacancies and consistent with a $p(2\times2)$ reconstruction. Three other orientations related by in-plane 90$^\circ$ rotations are also permitted. Upper left and lower right shaded squares shows the original unit cell and the doubled unit cell, respectively.  (b) Layer-by-layer view showing the proposed oxygen vacancy reconstruction of the terminal CuO$_2$ plane (highlighted in yellow). The vacancies result in a net $-1$ charge per unit cell on the topmost plane and a transfer of $-1$ charge per unit cell to the bottom of the film, avoiding the electric potential divergence associated with a polar catastrophe.}}
\end{figure}

The measured Sr$_{1-x}$La$_{x}$CuO$_{2}$ films are terminated with a CuO$_2$ layer, and the extra RHEED streaks associated with the reconstruction, shown in Fig.\ \ref{figure6}(b), become more intense during the vacuum annealing step that occurs at the end of film growth. This strongly suggests that the reconstruction is related to the removal of oxygen from the topmost CuO$_2$ plane. Indeed, half an oxygen vacancy per unit cell will change the net charge of the terminal CuO$_2$ atomic layer from $-2$ to $-1$, resulting in a cancellation of the divergent electric potential associated with alternating charged layers \cite{nakagawa2006}. Under this constraint, only one structure is consistent with a $p(2\times2)$ symmetry. One orientation of this structure is displayed in Fig.\ \ref{figure7}, with three others related by in-plane 90$^\circ$ rotations. The measured widths of both RHEED streaks and LEED spots indicate that such oxygen vacancy ordering persists over an in-plane length scale of at least $\sim$10 unit cells. We emphasize that the clear $(\pm1/2,\pm1/2)$ diffraction peaks in LEED measurements definitively rule out the possibility that the surface consists of domains of $p(2\times1)$ and $p(1\times2)$ reconstructions, a fact greatly limiting the set of possible oxygen vacancy structures to consider. If a single domain of our proposed model existed on the film surface, an anisotropic LEED structure factor would be observed. Because we instead observe four-fold symmetric diffraction patterns, the films likely contain domains of all four rotational orientations.

Higher doping levels lead to better metallic screening of the polar electric potential divergence, and the band gap of a material can play a large role in its polar reconstruction \cite{gu2009}. It is therefore natural to expect the tendency towards a polar surface reconstruction to be diminished at higher doping levels.  Indeed, we note that for $x \leq 0.05$, about half of films showed evidence of a reconstruction either by RHEED or LEED, while for $x \approx 0.10$, only one out of eight films showed the phenomenon. The observed film-to-film variability at fixed doping may be related to the fact that the formation of long-range structural order at the surface of the film, necessary for observing the reconstruction with diffraction probes, is likely sensitive to temperature, oxygen partial pressure, and other growth parameters.

\begin{figure}
\includegraphics{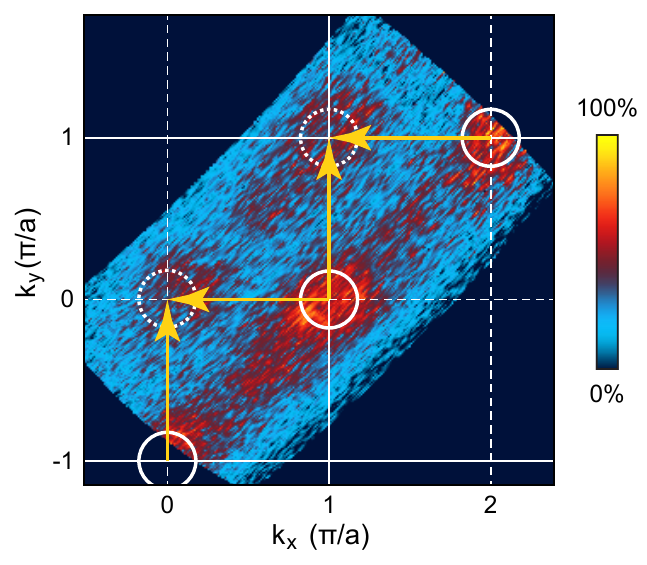}
\caption{\label{figure8}{Fermi surface map for a Sr$_{0.95}$La$_{0.05}$CuO$_{2}$ film taken at 20 K showing spectral weight within $E_F \pm 50$ meV and normalized to a featureless background at high binding energy. The Fermi surface is composed of small electron pockets centered at $(\pi,0)$ and equivalent points. Weak spectral weight visible at $(0,0)$ and $(\pi,\pi)$ is a result of shadow band reflections (shown by yellow arrows) due to a $p(2\times2)$ surface reconstruction in this film.}}
\end{figure}

Despite a dramatic reconstruction of the terminal CuO$_2$ surface in Sr$_{1-x}$La$_{x}$CuO$_{2}$, ARPES measurements are mostly consistent with a pristine unreconstructed material, with the Fermi surface enclosing the correct Luttinger volume. Figure \ref{figure8} shows a Fermi surface map for an $x = 0.05$ sample, identical to the top panel of Fig.\ \ref{figure2}(b), where there is some evidence of a $p(2\times2)$ reconstruction in the form of weak shadow bands at $(0,0)$ and $(\pi,\pi)$. With this exception, the ARPES data can be analyzed without considering the reconstruction. If the model presented above is correct, the large number of oxygen vacancies on the film surface will naturally alter the valence states of the topmost copper and oxygen atoms, moving their energies away from the Fermi level. Low-energy photoemission on such films will therefore effectively probe the first buried CuO$_2$ plane in the material. This will slightly suppress the photoemission intensity near the Fermi level because photoelectrons must travel through the top atomic layer before leaving the sample, explaining the high relative background intensity that is observed in the photoemission data of all Sr$_{1-x}$La$_{x}$CuO$_{2}$ films studied.

\begin{figure}
\includegraphics{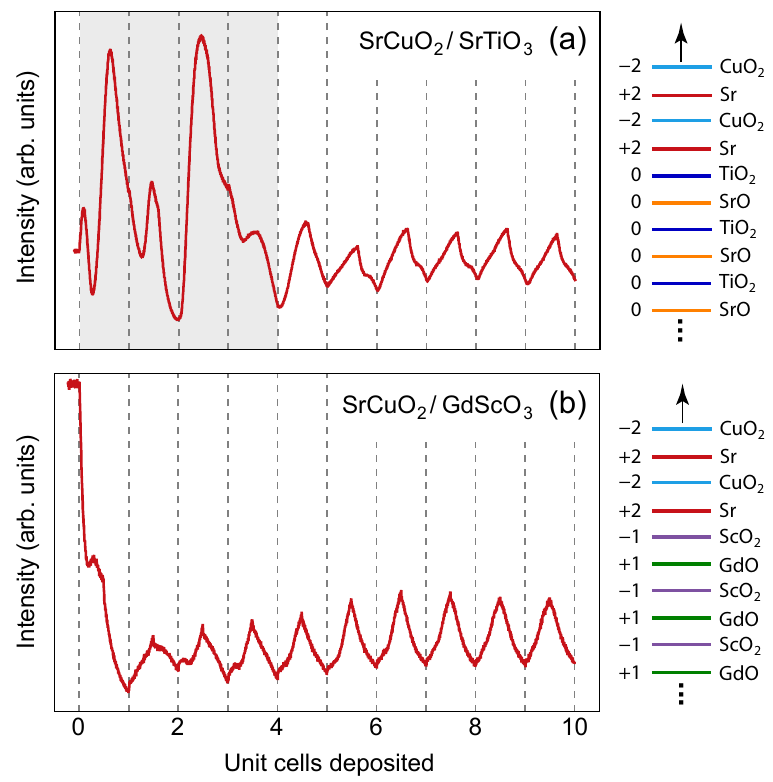}
\caption{\label{figure9}{Oscillations of the intensity of the [10] diffraction rod (RHEED oscillations) at the start of film growth.  (a) Thickness-controlled transition for a SrCuO$_2$ film grown on nonpolar (001) SrTiO$_3$. Deposition of the first four unit cells results in a dramatically varying RHEED intensity (highlighted in gray). Stable oscillations are observed only after depositing the fourth unit cell.  (b) Stable oscillations for a film grown on (110) GdScO$_3$. Diagrams at right illustrate individual atomic layers for a two unit cell thick film, with arrows representing the growth direction.}}
\end{figure}

Recently, Ref.\ \citenum{zhong2012} predicted that ultrathin films of polar SrCuO$_2$ grown on nonpolar SrTiO$_3$ substrates would exhibit a thickness-controlled transition from a chain-type structure for $\le4$ unit cells to the infinite-layer structure for $\ge5$ unit cells. As Fig.\ \ref{figure9} shows, we observe evidence for such a transition in RHEED oscillations during the growth of the first few unit cells of SrCuO$_2$ on (001) SrTiO$_3$. The first four unit cells consistently show a different pattern of oscillations, and only after deposition of the fourth unit cell do the typical oscillations begin. Interestingly, when films are grown on polar (110) GdScO$_3$ substrates, this behavior is suppressed and the first four deposition periods show oscillations qualitatively similar to those at later times. This atomic reconstruction, occurring during the formation of the first few unit cells of SrCuO$_2$, is likely related to electrostatic instabilities of both the film and the substrate. This may explain why film growth on a polar substrate such as GdScO$_3$ results in qualitatively different RHEED oscillations even during deposition of the first atomic layers.

\section{Conclusion}
\label{sectionConclusion}

In conclusion, we have used \textit{in situ} ARPES in conjunction with molecular-beam epitaxy to study the doping evolution of thin films of the infinite-layer electron doped cuprate Sr$_{1-x}$La$_{x}$CuO$_{2}$. At low doping, a dispersive LHB characteristic of cuprate parent compounds is observed. With the addition of electron carriers, a transition from charge-transfer insulator to metallic superconductor is observed as spectral weight, first appearing at $(\pi,0)$, gradually fills in the charge-transfer gap. Most notably, we observe the coexistence of a remnant LHB with the coherent low-energy states, even for $x = 0.10$. Electron diffraction was used to study the $p(2\times2)$ polar surface reconstruction observed in low-doped films. This reconstruction can be explained with a simple model by considering the polar nature of Sr$_{1-x}$La$_{x}$CuO$_{2}$. Finally, we have presented evidence supporting the theoretical prediction of a thickness-controlled transition in ultrathin films of Sr$_{1-x}$La$_{x}$CuO$_{2}$ grown on nonpolar substrates. Our work shows that strong local correlations, forming a remnant LHB, remain important in the cuprates even at high electron-doping levels. Furthermore, it appears that in films of Sr$_{1-x}$La$_{x}$CuO$_{2}$ in the ionic limit at low doping, a polar catastrophe is avoided via a structural rather than an electronic reconstruction. Our observations highlight the need for deeper investigation into the stability and structural changes that occur in polar complex oxide thin films and their surfaces.

\section*{ACKNOWLEDGMENTS}

We are grateful to Haofei Wei for help with sample characterization. This work was supported by the Air Force Office of Scientific Research (Grant No.\ FA9550-11-1-0033) and the National Science Foundation through the MRSEC program (Grant No.\ DMR-1120296). L.M.\ was supported by the Army Research Office (Grant No.\ W911NF-09-1-0415), D. E. S. acknowledges support from the National Science Foundation under Grant No.\ DGE-0707428 and through the IGERT program under Grant No.\ DGE-0654193, and E. J. M. acknowledges support from an NSERC PGS.

\end{document}